\documentclass[useAMS,usenatbib]{mn2e}

\usepackage{graphicx}

\title[Companions to X-ray emitting late B-type stars]{
Establishing the nature of companion candidates to X-ray emitting late B-type stars\thanks{Based on observations obtained at the European Southern
Observatory, Paranal, Chile (ESO programmes 074.D-0374(A) and 075.C-0522(A)).}}
\author[S. Hubrig et al.]{S. Hubrig$^{1}$\thanks{E-mail: shubrig@eso.org},
O. Marco$^{1}$, B. Stelzer$^{2}$, M. Sch\"oller$^{1}$, N. Hu\'elamo$^{3}$\\ 
$^{1}$European Southern Observatory, Casilla 19001, Santiago, Chile\\
$^{2}$INAF -- Osservatorio Astronomico di Palermo, Piazza del Parlamento 1, 90134 Palermo, Italy\\
$^{3}$Laboratorio de Astrofisica Espacial y Fisica Fundamental, LAEFF-INTA, Madrid, Spain\\
}

\begin{document}

\date{Accepted 2007 Enero 99. Received 2007 Enero 98}

\pagerange{\pageref{firstpage}--\pageref{lastpage}} \pubyear{2006}

\maketitle

\label{firstpage}

\begin{abstract}
The most favored interpretation for the detection of X-ray emission from late B-type 
stars is that these stars have a yet undiscovered late-type
companion (or an unbound nearby late-type star) that produces the
X-rays. Several faint IR objects at (sub)-arcsecond separation from
B-type stars have been uncovered in our earlier adaptive
optics imaging observations, and some of them have been followed up
with the high spatial resolution of the {\em Chandra} X-ray observatory,
pinpointing the X-ray emitter.
However, firm conclusions on their nature requires a search for
spectroscopic signatures of youth. Here we report on
our recent ISAAC observations carried out in low resolution spectroscopic mode. Equivalent 
widths 
have been used to obtain information on 
spectral types of the companions. All eight X-ray emitting systems with late B-type 
primaries studied contain dwarf like companions with spectral types later than A7.
The only system in the sample where the companion turns out to be
of early spectral type is not an X-ray source. These results are 
consistent with the assumption
that the observed X-ray emission from late B-type stars is produced by an active 
pre-main sequence companion star. 
\end{abstract}

\begin{keywords}
binaries: visual -
binaries: spectroscopic -
stars: pre-main-sequence -
X-rays: stars
\end{keywords}

\section{Introduction}
\label{sect:intro}

X-ray observations performed with the {\em Einstein Observatory} and {\em ROSAT}
missions have revealed that X-rays are emitted by stars throughout the
Hertzsprung-Russell diagram.  For stars on the main-sequence (MS) two
mechanisms are known to be responsible for the observed emission: In O-
and early B-type stars, the X-rays are produced by instabilities
arising in their strong radiatively driven stellar winds
\citep{1999ApJ...520..833O,1980ApJ...241..300L},
and in late-type stars a solar-like
magnetic dynamo is thought to be responsible for the observed X-ray activity
\citep{1955ApJ...122..293P}.
No X-ray emission is expected from stars whose spectral types are
late~B and early~A.
Nevertheless, X-ray detections of these
stars have been reported in several works
\citep[e.g.,][and references therein]{2002ApJ...578..486D}.

In absence of another explanation, the X-ray emission of MS late B-type and
early A-type stars is commonly attributed to unresolved late-type companions.
During the last years, in order to obtain clarification of this issue, we pursued a multi-fold strategy:
(i)~an Adaptive Optics search for previously unresolved late-type
companions to B- and A-type stars which are known to be X-ray emitters,
(ii)~high spatial resolution follow-up X-ray observations
of those with new companions to identify the X-ray source within these systems,
and (iii)~low-resolution spectroscopy of the X-ray emitting objects identified
with Adaptive Optics to establish their nature.

In 1999 we have been engaged in a multiplicity survey of 
{\em ROSAT} selected bright late B-type dwarfs using the European Southern Observatory's
ADONIS (Adaptive Optics Near Infrared System) instrument,
searching for close companions in the diffraction-limited IR images
\citep{2001A&A...372..152H}. 
For the systems studied, the X-ray flux levels observed in the {\em ROSAT} All-Sky Survey (RASS)  
are $\log{L_{\rm X}}\,{\rm [erg/s]} \sim 29.5-30.5$ \citep{1996A&AS..118..481B},
far too high to be explained by emission from cooler MS companion stars in the system. 
Nearby late-type field stars display much lower activity levels,
with typically $\log{L_{\rm X}}\,{\rm [erg/s]} \sim 26-28$ \citep{2004A&A...417..651S}.
The observed X-ray luminosities are similar to those of T\,Tauri stars, that show 
typically $\log{L_{\rm X}}\,{\rm [erg/s]} \sim 28-31$ \citep{2005ApJS..160..401P}.
Therefore, a reasonable interpretation within the companion scenario is 
that the X-ray emission is produced by active pre-main sequence (PMS) companion stars.
Since the late B-type stars that we have studied are rather young with ages
less than a few hundred million years \citep{2001A&A...372..152H},
the detected new low-mass companion candidates are, indeed, expected to be 
PMS stars or very young MS stars. 

In the sample of 49 late B-type stars observed with ADONIS, we found 29 faint objects
near 25 stars.  For 21 of these 29 IR objects 
our ADONIS images represent a first detection. 
If all these objects are true physical companions, 
the resulting binary frequency in this spectral type domain is 51\%.
However, our sample was biased towards
low-mass companions that exhibit strong X-ray emission.
For the discovered IR objects the following information has
been collected in the previous studies:
For 11 of the faint IR objects both J and K magnitudes were obtained with ADONIS. 
The near-IR photometry of these objects was 
compared to the evolutionary models for low-mass
PMS stars calculated by \citet{1998A&A...337..403B}.
For seven of them, Fig.~2 of \citet{2001A&A...372..152H} shows that their position 
is compatible with them being low-mass PMS stars.
In the remaining four systems for which we obtained both J and K measurements with ADONIS
the discovered IR objects have already evolved to the MS,  
and could not be fitted with PMS evolutionary tracks.
For the companion candidates in the other $14$ ADONIS systems 
to date only K-band measurements are available. 
Knowing the age of the late B-type primaries from their position in the H-R diagram and assuming that 
systems are coeval, we placed the companions with known $M_k$ values along the isochrones 
of the B primaries to estimate their masses, luminosities and effective temperatures \citep{2001A&A...372..152H}.
From these considerations we expect that out of these
14~systems 9~contain PMS companions. 
Thus, our photometric study suggested that out of the observed
IR objects, 16~are PMS stars. 
Their masses estimated from the PMS tracks range from about 1.2\,M$_\odot$ down to 0.6\,M$_\odot$.

In the framework of our accompanying {\em Chandra} program we have obtained
high spatial resolution ($\sim$1\arcsec{}) X-ray observations of late B-type stars
with companion candidates from near-IR photometry. Contrary to the earlier RASS data,
in the {\em Chandra} images we were able to 
separate the contributions of the B-type star and the known visual companions to the X-ray emission. 
The {\em Chandra} sample was composed 
of $11$ systems with late B-type primaries and $15$ known visual companions, most of them
identified in our ADONIS survey. 
We found X-ray emission from $12$ of the new IR objects; in one case the companion is
a binary that can not be resolved with {\em Chandra} \citep{2003A&A...407.1067S,2006A&A...452.1001S}.
The companion candidates detected with {\em Chandra} show X-ray luminosities in the range
$\log{L_{\rm X}}\,{\rm [erg/s]} \sim 29-30$,
typical for T\,Tauri stars \citep[e.g.,][]{2005ApJS..160..401P}.
These objects can, therefore, be considered as strong candidate PMS stars. 

However, companionship can not be established based on photometry and X-ray
observations alone.
To confirm or reject the candidates identified as described above, 
we carried out a spatially resolved K-band spectroscopic study of the companions 
with ISAAC at the VLT. Our objective was to find out whether the  
detected IR- and X-ray candidate PMS companions are physically associated objects or 
background sources. In addition, for three studied systems we present our recent observations 
carried out with the
Nasmyth Adaptive Optics System with Near-Infrared Imager and Spectrograph (NACO) at the VLT.

\section{Sample} 

Our sample of eight systems selected for ISAAC observations is based on the photometric
survey with ADONIS and contains companions which can be easily 
separated with ISAAC from their parent stars during good seeing conditions and have been visible 
during our visitor run on May~24, 2005.
Four of these systems were in the {\em Chandra} sample, and all their companions were detected
at levels of $\log{L_{\rm X}}\,{\rm [erg/s]} \sim 29-30$ \citep{2006A&A...452.1001S}.

In addition, we observed HD\,165493 and HD\,104237-6. 
The system HD\,165493 was previously studied by \citet{1985A&AS...60..183L} 
in the course of his survey of visual double stars with early-type primaries.
\citet{1996A&AS..118..481B} failed 
to detect any X-ray emission from this system in the RASS (the companion at $4^{\prime\prime}$
was not resolved). However, the upper limit for the X-ray flux of 
$\log{L_{\rm X}}\,{\rm [erg/s]} < 31.4$ 
is quite high and does not rule out the presence of X-ray emission
at typical T\,Tauri star levels. 
The T\,Tauri star HD\,104237-6 is a low-mass companion to the optically brightest 
Herbig Ae star HD\,104237 at a separation of 14\farcs{}88 \citep{2004ApJ...608..809G}.
It was used in our observations as a comparison star.
As was shown by \citet{2003ApJ...599.1207F} and \citet{2004ApJ...608..809G},
HD\,104237-6 exhibits H$\alpha$ emission at a level typical of weak-line T\,Tauri stars. 
The X-ray properties of this star 
are also T\,Tauri like, with $\log{L_{\rm X}}\,{\rm [erg/s]} = 30.7$ according to \citet{2006A&A...457..223S}.

\begin{table*}
\caption{
Astrometry of late B-type stars with companions from ISAAC (cols.\ 2--5), ADONIS (cols.\ 6--8) and NACO (cols.\ 9/10).
The errors for the projected separation measurements are
$\pm$0\farcs{}07 for ISAAC, 
$\pm$0\farcs{}05 for ADONIS, and
$\pm$0\farcs{}02 for NACO.
The errors for the position angle are
$\pm 2^{\circ}$ for ISAAC,
$\pm0.2^{\circ}$ for ADONIS,
and $\pm0.1^{\circ}$ for NACO.
}
\begin{center}
\begin{tabular}{lccrccccrrccr}
\hline
\hline
\multicolumn{1}{c}{Object} &
\multicolumn{1}{c}{\hfil} &
\multicolumn{1}{c}{Separation} &
\multicolumn{1}{c}{PA} &
\multicolumn{1}{c}{Modified} &
\multicolumn{1}{c}{S/N} &
\multicolumn{1}{c}{\hfil} &
\multicolumn{1}{c}{Separation} &
\multicolumn{1}{c}{PA} &
\multicolumn{1}{c}{$\Delta{}K$} &
\multicolumn{1}{c}{\hfil} &
\multicolumn{1}{c}{Separation} &
\multicolumn{1}{c}{PA} \\
\multicolumn{1}{c}{} &
\multicolumn{1}{c}{} &
\multicolumn{1}{c}{[\arcsec]} &
\multicolumn{1}{c}{} &
\multicolumn{1}{c}{Julian date} &
\multicolumn{1}{c}{} &
\multicolumn{1}{c}{} &
\multicolumn{1}{c}{[\arcsec]} &
\multicolumn{1}{c}{} &
\multicolumn{1}{c}{} &
\multicolumn{1}{c}{} &
\multicolumn{1}{c}{[\arcsec]} &
\multicolumn{1}{c}{} \\
\multicolumn{1}{c}{} &
\multicolumn{1}{c}{} &
\multicolumn{4}{c}{ISAAC} &
\multicolumn{1}{c}{} &
\multicolumn{3}{c}{ADONIS} &
\multicolumn{1}{c}{} &
\multicolumn{2}{c}{NACO} \\
\hline
HD\,73340 & & 0.5 & 214 & 53516.022 & 220 & & 0.604 & 221.2 & 2.52 & & 0.570 & 220.0 \\
HD\,73952 & & 1.0 & 207 & 53516.040 & 180 & & 1.162 & 205.3 & 4.19 & & \\
HD\,78556 & & 1.3& 299 & 53516.060 & 210 & & 1.300 & 298.5 & 3.31 & & \\
HD\,110073 & & 1.1 & 74 & 53516.101 & 260 & & 1.192 & 75.0 & 3.08 & & 1.202 & 73.9 \\
HD\,114911 & & 2.6 & 124 & 53516.147 & 130 & & 2.706 & 124.6 & 0.319 & & \\
HD\,133880 & & 1.2 & 111 & 53516.117 & 150 & & 1.222 & 109.2 & 2.35 & & \\
HD\,134837 & & 4.7 & 155 & 53516.162 & 120 & & 4.696 & 154.3 & $\ge$4.66 & & \\
HD\,165493 & & 4.1 & 258 & 53516.179 & 360 & & & & & & 4.043 & 257.3\\
HD\,184707 & & 2.4 & 175 & 53516.194 & 150 & & 2.435 & 173.1 & $\ge$2.76 & & \\
HD\,104237-6 & & & & 53516.134 & 180 \\
\hline
\end{tabular}
\end{center}
\label{tab:sources}
\end{table*}

\section{Observations and data reduction}
\label{sect:obs}

The observations were carried out 
using the ISAAC infrared spectrograph on UT1 at the VLT in low-resolution mode (R=1500)
in clear weather conditions with seeing around 0.8 to 1.0\arcsec.
The pixel size was 0.147\arcsec{}/pixel with a slit size of 0.3\arcsec{}.
The exposure time for each target has been adequately chosen to reach a good S/N ($>$100).
The targets and telluric standards were observed with the traditional long-slit nodding pattern 
with a few arcsecond jitter to better sample the detector.
A telluric standard star was observed either just before or after the target object. 
The acquisition images for each target and slit position are presented in Fig.~\ref{fig:acq}. 
A flat-field correction has been applied to all frames, using a flat-field lamp for internal calibration.
The background has been removed using pairs of frames, followed by dead pixel removal.
The wavelength calibration model has been obtained using Xenon and Argon lamp arc observations.
The extracted target spectra have been corrected for the telluric features using typically two 
standard stars of late B- and early A-dwarf spectral type. We also obtained spectra of a few early-G dwarfs.
A set of these standard stars of solar spectral type has been used to trace possible features
produced by $Br\gamma$ absorptions visible in the hot standard stars and to correct for them.
The final normalized spectra of the studied companions are presented in Fig.~\ref{fig:spectra}.
We note that very faint emission features at 2.317\,$\mu$m in the spectra of HD\,73340 
and HD\,110073 are reduction artifacts left after the correction for terrestrial absorption.

The system HD\,184707 is a possible triple system: we discover just at the detection limit a very 
faint  additional companion  on the 
acquisition image.
It is located at 1.3\arcsec{} from the primary star at a position angle of 73$^\circ$
(see Fig.~\ref{fig:acq}).
However, the presence of this companion has still to be confirmed by deeper observations.

\section{Results}
\label{sect:res}

\begin{figure}
\centering
\includegraphics[width=0.45\textwidth]{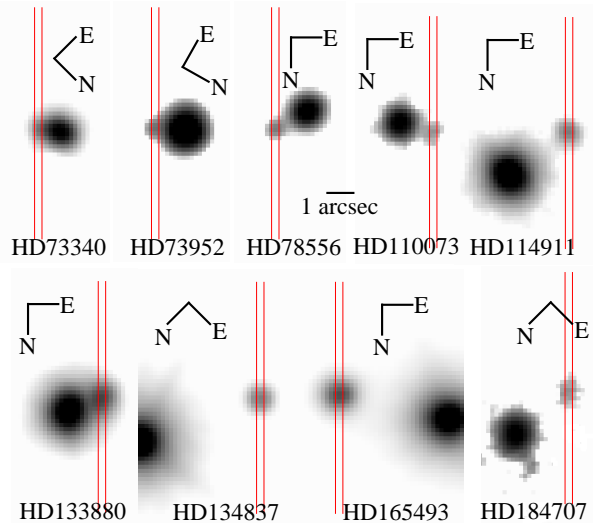}
\caption{
Acquisition images and slit positions for the observed systems with companions.
}
\label{fig:acq}
\end{figure}

\begin{figure}
\centering
\includegraphics[width=0.45\textwidth]{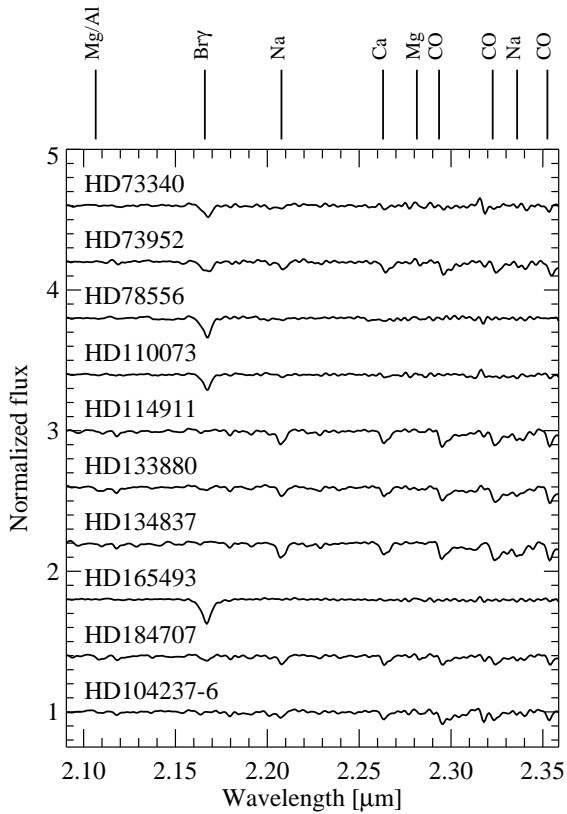}
\caption{
Normalized ISAAC spectra of the studied companions. All but the lowest are displaced upwards 
for display purposes.
}
\label{fig:spectra}
\end{figure}

In Table~\ref{tab:sources} we present the astrometry of the 
X-ray selected late B-type stars with companions. 
The columns from left to right give the name of the object,
the separation, position angle (PA), modified Julian date, and signal-to-noise ratio (S/N)
of the final one-dimensional spectra for the new ISAAC measurements,
the separation, position angle, and binary magnitude difference ($\Delta{}K$) from our 1999 ADONIS observations \citep{2001A&A...372..152H},
and finally the separation and position angle from our recent NACO observations at the VLT obtained in February and March 2005
(\citealt{2005astro.ph.10302H}; Sch\"oller et al.\ (in prep.)).
The separation and position angles for ISAAC were measured on the acquisition images presented in Fig.~\ref{fig:acq}.
The accuracy of the projected separation measurements in ISAAC acquisition images is about $\pm$0\farcs{}07, and 
the accuracy of the PA is typically about $\pm 2^{\circ}$.
The uncertainties of the projected separation measurements for ADONIS and NACO are 
about $\pm$0\farcs{}05 and $\pm$0\farcs{}02, respectively.
For the position angles, the accuracies are 
$\pm0.2^{\circ}$ for ADONIS and $\pm0.1^{\circ}$ for NACO.
We have not attempted to make any astrometric study, since the data for the different epochs were obtained
with different instruments.
However, the values for separation and 
position angles presented in Table~\ref{tab:sources} show a good consistency of the position 
of the companions relative to the primary stars from 1999 (ADONIS observations) to 2005 (ISAAC and NACO observations)
and did not change much within six years. 

\subsection{Spectral classification}
\label{sect:classif}

\begin{table}
\caption{The strongest features in the K-band spectra.
}
\begin{center}
\begin{tabular}{cc}
\hline
\hline
\multicolumn{1}{c}{Feature} &
\multicolumn{1}{c}{$\lambda$ [$\mu$m]} \\
\hline
Br$\gamma$ & 2.16 \\
Na~I & 2.20, 2.21 \\
Ca~I & 2.26 \\
Mg~I & 2.28 \\
CO & 2.29, 2.32, 2.35 \\
\hline
\end{tabular}
\end{center}
\label{tab:features}
\end{table}

The wavelengths of the strongest features in our K-band spectra are listed in Table 2. 
As was shown by \citet{1998ApJ...497..354L}, the metallic absorption features can be used in the spectral 
classification for spectral types later than F8.
The only strong feature expected in the K-band spectrum of the late B-type primaries is Br$\gamma$.
 
The equivalent widths (EW) of all features 
were measured by defining a global continuum as the best-fit line to wavelength regions which are free 
of spectral lines. 
The absorption features presented in Table~\ref{tab:features} are sensitive to temperature and their equivalent 
widths have been used for spectral classification as suggested by \citet{1998ApJ...497..354L}.

For very close faint companions, some contamination of the spectra by the primary stars is expected, 
but this will not affect the Na, Ca, Mg, and CO features.
In hotter and more massive companion candidates with spectral types from A0 to F5 the strongest 
K-band feature is Br$\gamma$.
In these cases we used the atlases by \citet{1992A&AS...96..593L} and \citet{1997ApJS..111..445W} which present 
K-band spectra of  stars of different spectral types (O to M) and luminosity types I to V. 
The equivalent widths of the K-band absorption features in the spectra of all companions and the comparison 
star HD\,104237-6 are presented in Table~\ref{tab:values}. The corresponding spectral types are listed in the last 
column.

\subsection{Brief notes on individual targets}
\label{sect:stars}

In the following paragraphs we compare our findings from our previous near-IR study with 
the results of the K-band spectroscopy.
%

\begin{table*}
\caption{
Equivalent widths of the K-band absorption features.
}
\begin{center}
\begin{tabular}{lcccccc}
\hline
\hline
\multicolumn{1}{c}{Object} &
\multicolumn{5}{c}{Equivalent Width [\AA{}]} &
\multicolumn{1}{c}{Spectral Type} \\
\multicolumn{1}{c}{} &
\multicolumn{1}{c}{Br$\gamma$} &
\multicolumn{1}{c}{Na\,I} &
\multicolumn{1}{c}{Ca\,I} &
\multicolumn{1}{c}{Mg\,I} &
\multicolumn{1}{c}{$^{12}$CO(2,0)} &
\multicolumn{1}{c}{} \\
\hline
HD\,73340    & 4.5$\pm$0.2 & 1.8$\pm$0.2  &0.7$\pm$0.1 &0.8$\pm$0.1   & 0.1$\pm$0.1  & F5-F7V  \\
HD\,73952    & 4.2$\pm$0.2 & 3.8$\pm$0.2 & 4.0$\pm$0.1 & 1.3$\pm$0.1 & 7.9$\pm$0.1 & K6-M0V  \\
HD\,78556    &7.1$\pm$0.2 & 0.3$\pm$0.1 & 0.2$\pm$0.1  &0.2$\pm$0.1   &0.1$\pm$0.1   & A7-A8V  \\
HD\,110073   & 5.4$\pm$0.2 &0.6$\pm$0.1  & 0.5$\pm$0.1  & 0.3$\pm$0.1   &0.1$\pm$0.1   & F4-F5V  \\
HD\,114911   &0.4$\pm$0.2 & 4.9$\pm$0.2 & 4.2$\pm$0.1  &0.9$\pm$0.1  & 9.0$\pm$0.1  &K6-M0V   \\
HD\,133880   &1.6$\pm$0.2 &2.8$\pm$0.2  &2.8$\pm$0.1 &1.2$\pm$0.1   & 7.0$\pm$0.1  &  K2-K3V \\
HD\,134837   &0.4$\pm$0.2 &5.0$\pm$0.2  &4.2$\pm$0.1   &0.5$\pm$0.2   & 10.1$\pm$0.2  & K6-M0V  \\
HD\,165493   &11.1$\pm$0.2 &0.0$\pm$0.1 &0.0$\pm$0.1 &0.0$\pm$0.1 &0.0$\pm$0.1 & A3-A5V  \\
HD\,184707   &1.5$\pm$0.2 &3.0$\pm$0.2   &3.1$\pm$0.2  & 1.3$\pm$0.1  & 4.6$\pm$0.1  & K2-K4V  \\
HD\,104237-6 &1.0$\pm$0.2 & 3.1$\pm$0.2  &2.6$\pm$0.1   &1.2$\pm$0.1   & 6.8$\pm$0.1  & K3-K4V  \\
\hline
\end{tabular}
\end{center}
\label{tab:values}
\end{table*}


\subsubsection{The comparison star HD\,104237-6}
\label{sect:104237}

This star, which is a low-mass companion to the optically brightest Herbig Ae star HD\,104237 
\citep{2004ApJ...608..809G}, has a spectral type K3\,IV  determined from 
optical spectroscopy with FEROS.
We use this object as a comparison star.
The equivalent widths of the absorption features in our K-band 
spectrum indicate a spectral classification K3-K4. 
The strength of CO relative to Na and Ca is dwarf-like in agreement with the determination of 
\citet{2004ApJ...608..809G}.

\subsubsection{HD\,73340}
\label{sect:73340}

The PMS companion with a mass of 1.2\,M$_\odot$ was discovered with the European Southern 
Observatory's ADONIS instrument in March 1999 \citep{2001A&A...372..152H} and later confirmed by 
observations with NACO at the VLT in February 2005 (Sch\"oller et al.\ (in prep.)). 
The strongest spectral feature is the Br$\gamma$ absorption line indicating a spectral type F5-F7 
according to the near-IR stellar atlases of \citet{1992A&AS...96..593L} and \citet{1997ApJS..111..445W}. 
The Na, Ca, Mg, and CO features are all weak, 
in agreement with a spectral type of a companion of 
a mass of 1.2\,M$_\odot$. 

\subsubsection{HD\,73952}
\label{sect:73952}

\citet{2001A&A...372..152H}
suggested that the mass of the companion is about 0.6\,M$_\odot$ 
Using equivalent widths of  Na, Ca, Mg, and CO spectral features 
we conclude that the companion is a dwarf with a spectral type K6--M0. 
The Br$\gamma$ line appears much too strong for the determined spectral type.
However, because of the faintness of the companion ($\Delta{}K$ = 4.19; cf.\ Table~\ref{tab:sources}) and its closeness to the 
primary (see Fig.~\ref{fig:acq}), the 
primary clearly  contributes to the K-band spectrum, giving rise to the composite spectrum we observe.  

\subsubsection{HD\,78556}
\label{sect:78556}

This star was observed with ADONIS only in the K-band and the determination of the mass of the companion
was very uncertain.
The Br$\gamma$ strength and the weakness of other 
absorption features constrain the spectral type to about A7-A8 according to the 
near-IR stellar atlases.
\citet{1996A&AS..118..481B} measured an X-ray luminosity of $\log{L_{\rm X}}\,{\rm [erg/s]} \sim 29.7$.

\subsubsection{HD\,110073}
\label{sect:110073}

According to our previous study, the companion is a PMS star with a mass of 1.13\,M$_\odot$. 
The K-band spectrum appears similar to that of the companion to HD\,73340 with very faint Na, Ca, Mg,
and CO features and a rather strong Br$\gamma$ line. Their 
strengths  match with the spectra presented in the stellar atlases of a spectral type of F4-F5. 

\subsubsection{HD\,114911}
\label{sect:114911}

The mass of the PMS companion was estimated as 0.88\,M$_\odot$. The Br$\gamma$ absorption is very weak, 
and the strength of other features implies a spectral type between K6 and M0. 

\subsubsection{HD\,133880}
\label{sect:133880}

According to our previous study, the companion to HD\,133880 is expected to be a  
PMS star with a mass of 1.17\,M$_\odot$. The spectrum fits with a spectral type K2-K3.

\subsubsection{HD\,134837}
\label{sect:134837}

Only K-band imaging was done for this system in our previous work.
The K-band spectrum of the companion resembles rather well the spectrum of the 
companion to HD\,114911. From the strength of Na, Ca, Mg, and CO we derive a spectral type between  K6 and M0.

\subsubsection{HD\,165493}
\label{sect:165493}

Using  K-band imaging
carried out with NACO in March 2005, we detected a companion at a separation of about 4$\arcsec$ with
a K-magnitude about 3\,mag fainter than the primary star.
The spectrum exhibits only strong  Br$\gamma$ 
absorption and no other features are detected. The strength of the Br$\gamma$ line implies an early 
spectral type between A3 and A5.

\subsubsection{HD\,184707}
\label{sect:184707}

Only K-band imaging was done for this system in our previous work.
The equivalent widths of absorption features in our K-band 
spectrum imply a spectral classification K2--K4.

\section{Discussion}
\label{sect:disc}

The detection of X-ray emission from late B- and early A- type stars has been reported from virtually 
all X-ray satellites and has long remained a puzzle. MS stars 
of spectral type $\sim$B2 to A7 do not 
show high-speed stellar winds nor deep convection zones needed to support dynamo 
action; hence no X-ray emission is predicted for them. 
It has long been argued that late-type companion stars might be responsible for
the observed X-rays but this hypothesis could not be tested observationally with previous
low-spatial resolution instruments. In the past we have approached the problem with 
high-spatial resolution imaging in the near-IR with ADONIS and in X-rays with {\em Chandra}. The results
from these studies suggest that many late B-type stars drawn from 
the {\em ROSAT} All-Sky Survey (RASS) have faint IR objects nearby and these companion candidates 
are responsible for at least a fraction of the X-ray emission.
 
To examine the nature of these objects we performed follow-up spectroscopy.  
In the sample of nine candidate multiple systems presented here we confirm the PMS nature of
the companion candidate. 
In five systems (HD\,73953, HD\,114911, HD\,133880, HD\,134837, and HD\,184707) the studied companions 
exhibit dwarf-like K-band spectra consistent with spectral types
between K and M. The spectra of these low-mass companions resemble well the spectrum of the 
T\,Tauri star HD 104237-6 of spectral type 
K3\,IV. All of them lack emission features at the positions
of the Br$\gamma$ line and of the CO band-heads, 
indicating that these companions could be classified as weak-line T\,Tauri stars.  
The companions of three systems (HD\,73340, HD\,78556, and HD\,110073)
are more massive and their K-band spectra are consistent with spectral types late A to mid F.
Considering that the spectra of all the studied companions 
suggest spectral types later than A7 and all of them show dwarf-like values for the luminosity indicator,
it is very unlikely that they are background sources.

In the case of HD\,165493, the companion in the system is relatively hot with a spectral type between 
A3 and A5. 
A detection of X-ray emission is not expected for stars of this spectral type because 
there is no known X-ray production mechanism. In fact, \citet{1996A&AS..118..481B} were
not able to detect X-ray emission from this system. However, their upper limit for the X-ray 
flux of $\log{L_{\rm X}}\,{\rm [erg/s]} < 31.4$ is not conclusive, and a deeper X-ray image
with {\em Chandra} resolving the binary would be useful. 

We note in this context, that
two more systems with suspected PMS companions discovered previously with ADONIS, 
HD\,75333 and HD\,145483, have been subsequently observed in 2001 with 
the NIRSPEC instrument at medium resolution R=3500 on the Keck II telescope together
with the AO system \citep{2005hris.conf..499H}. Similar to the results presented in this study, these 
observations have confirmed the late-type spectral classification with spectral types between K5 and M0.

Based on the results presented in this study we conclude that 
it is reasonable to attribute the observed X-ray emission from late B-type stars 
to active late-type PMS stars.  
The physical association of the studied companions with the B-type star 
is also anticipated from the consistency of our photometric observations with ADONIS 
with the predictions obtained from evolutionary models \citep{2001A&A...372..152H} 
as well as spatially resolved
X-ray images of a small sample carried out with {\em Chandra} \citep{2003A&A...407.1067S,2006A&A...452.1001S}.

Binary star formation mechanisms represent an important, but still not completely
understood part of star formation.
Considerable data have been accumulated for both MS and evolved 
late-type stars \citep{1991A&A...248..485D},
for PMS stars and for stars of intermediate mass and age in open clusters
\citep{1999A&A...341..547D,2004A&A...427..651D,2004AJ....127.1747H}.
Generally, these surveys come to the
conclusion that for the examined range of separations 
at young ages there are twice as many binaries as in the field.
In contrast, the multiplicity properties of higher-mass stars remain poorly known
and previous studies have provided inconclusive results.
The percentage of close (both visual and spectroscopic) binaries 
among B-type stars was found to be higher than among solar type stars,
and earlier spectral types seem to have a higher companion star fraction
than later spectral types  
(e.g.\ review by \citealt{2002ASPC..267..209Z}).
On the other hand, an adaptive optics survey in the ScoOB2 association led to  
the discovery of several new binaries of late B and A spectral type,
suggesting that the apparent decrease of the binary fraction towards 
later spectral types may be the result of observational biases 
\citep{2005A&A...430..137K}.
Searches for binaries among Herbig Ae/Be stars have been carried
out by means of IR imaging \citep{2001IAUS..200..155B}, optical
spectroscopy \citep{1999A&AS..136..429C}
and spectro-astrometry \citep{2006MNRAS.367..737B}.
All these studies suggested a high binary
frequency (the largest value, $68 \pm 11$\,\%, being achieved by 
the spectro-astrometric study), but were restricted to a limited sample.

Our previous ADONIS observations yielded an observed binary frequency of $51$\,\% for
the RASS selected sample. 
In this paper we have shown that, when followed-up
spectroscopically, most of the photometric companion candidates are confirmed 
as late-type stars. The high binary frequency in our sample
is certainly due to the fact that  the
stellar sample contained only X-ray selected stars and, in addition, was biased
towards
low-mass companions which exhibit strong X-ray emission. One of the future important
tasks would be to study a sample of late B-type stars not detected in the RASS.
Recently, Ivanov et al. (2006) have started a search of companions to 
intermediate-mass stars in the field. The main goal of their study is to 
derive the binary frequency of a volume-limited sample of BA-stars 
in the solar neighborhood with distances smaller than $300$\,pc. 
Such observations are necessary to get further insight in the formation mechanisms of
intermediate mass binaries.

\section*{Acknowledgments}

We would like to thank the anonymous referee for the useful comments that helped
us to improve this paper.


\label{lastpage}


\begin{thebibliography}{99}

\bibitem[\protect\citeauthoryear{Abt, Gomez, \& Levy}{1990}]{1990ApJS...74..551A}
Abt H.~A., Gomez A.~E., Levy S.~G., 1990,
ApJS, 74, 551 

\bibitem[\protect\citeauthoryear{Ali et al.}{1995}]{1995AJ....110.2415A} 
Ali B., Carr J.~S., Depoy D.~L., Frogel J.~A., Sellgren K., 1995,
AJ, 110, 2415 

\bibitem[\protect\citeauthoryear{Baines et al.}{2006}]{2006MNRAS.367..737B} 
Baines D., Oudmaijer R.~D., Porter J.~M., Pozzo M., 2006,
MNRAS, 367, 737 

\bibitem[\protect\citeauthoryear{Baraffe et al.}{1998}]{1998A&A...337..403B}
Baraffe I., Chabrier G., Allard F., Hauschildt P.~H., 1998,
A\&A, 337, 403 

\bibitem[\protect\citeauthoryear{Bergh\"ofer, Schmitt, \& Cassinelli}{1996}]{1996A&AS..118..481B}
Bergh\"ofer T.~W., Schmitt J.~H.~M.~M., Cassinelli J.~P., 1996,
A\&AS, 118, 481 

\bibitem[\protect\citeauthoryear{Bouvier \& Corporon}{2001}]{2001IAUS..200..155B}
Bouvier J., Corporon P., 2001,
IAUS, 200, 155 

\bibitem[\protect\citeauthoryear{Corporon \& Lagrange}{1999}]{1999A&AS..136..429C}
Corporon P., Lagrange A.-M., 1999, 
A\&AS, 136, 429

\bibitem[\protect\citeauthoryear{Daniel, Linsky, \& Gagn{\'e}}{2002}]{2002ApJ...578..486D}
Daniel K.~J., Linsky J.~L., Gagn{\'e} M., 2002,
ApJ, 578, 486 

\bibitem[\protect\citeauthoryear{Duch{\^e}ne}{1999}]{1999A&A...341..547D}
Duch{\^e}ne G., 1999,
A\&A, 341, 547 

\bibitem[\protect\citeauthoryear{Duch{\^e}ne et al.}{2004}]{2004A&A...427..651D}
Duch{\^e}ne G., Bouvier J., Bontemps S., Andr{\'e} P., Motte F., 2004,
A\&A, 427, 651

\bibitem[\protect\citeauthoryear{Duquennoy \& Mayor}{1991}]{1991A&A...248..485D}
Duquennoy A., Mayor M., 1991,
A\&A, 248, 485 


\bibitem[\protect\citeauthoryear{Feigelson, Lawson, \& Garmire}{2003}]{2003ApJ...599.1207F}
Feigelson E.~D., Lawson W.~A., Garmire G.~P., 2003,
ApJ, 599, 1207 

\bibitem[\protect\citeauthoryear{Grady et al.}{2004}]{2004ApJ...608..809G} 
Grady C.~A., et al., 2004,
ApJ, 608, 809 

\bibitem[\protect\citeauthoryear{Haisch et al.}{2004}]{2004AJ....127.1747H}
Haisch K.~E., Jr., Greene T.~P., Barsony M., Stahler S.~W., 2004,
AJ, 127, 1747 

\bibitem[\protect\citeauthoryear{Hubrig et al.}{2001}]{2001A&A...372..152H}
Hubrig S., Le Mignant D., North P., Krautter J., 2001,
A\&A, 372, 152 

\bibitem[\protect\citeauthoryear{Hubrig et al.}{2005}]{2005hris.conf..499H}
Hubrig S., Sch{\"o}ller M., Le Mignant D., Stelzer B., Hu{\'e}lamo N., Duch{\^e}ne G., 2005,
hris.conf, 499 

\bibitem[\protect\citeauthoryear{Hubrig, Ageorges, \& Sch\"oller}{2007}]{2005astro.ph.10302H}
Hubrig S., Ageorges N., Sch\"oller M., 2007,
[arXiv:astro-ph/0510302]

\bibitem[\protect\citeauthoryear{Ivanov et al.}{2006}]{2006Ap&SS.304..247I}
Ivanov V.~D., Chauvin G., Foellmi C., Hartung M., Hu{\'e}lamo N., Melo C., N{\"u}rnberger D., Sterzik M., 2006,
Ap\&SS, 304, 247 

\bibitem[\protect\citeauthoryear{Kouwenhoven et al.}{2005}]{2005A&A...430..137K}
Kouwenhoven M.~B.~N., Brown A.~G.~A., Zinnecker H., Kaper L., Portegies Zwart S.~F., 2005,
A\&A, 430, 137 

\bibitem[\protect\citeauthoryear{Lancon \& Rocca-Volmerange}{1992}]{1992A&AS...96..593L}
Lancon A., Rocca-Volmerange B., 1992,
A\&AS, 96, 593 

\bibitem[\protect\citeauthoryear{Leinert, Richichi, \& Haas}{1997}]{1997A&A...318..472L}
Leinert C., Richichi A., Haas M., 1997,
A\&A, 318, 472 

\bibitem[\protect\citeauthoryear{Lindroos}{1985}]{1985A&AS...60..183L} 
Lindroos K.~P., 1985,
A\&AS, 60, 183 

\bibitem[\protect\citeauthoryear{Lucy \& White}{1980}]{1980ApJ...241..300L}
Lucy L.~B., White R.~L., 1980,
ApJ, 241, 300 

\bibitem[\protect\citeauthoryear{Luhman \& Rieke}{1998}]{1998ApJ...497..354L}
Luhman K.~L., Rieke G.~H., 1998,
ApJ, 497, 354 


\bibitem[\protect\citeauthoryear{Morrell \& Levato}{1991}]{1991ApJS...75..965M}
Morrell N., Levato H., 1991,
ApJS, 75, 965 

\bibitem[\protect\citeauthoryear{Owocki \& Cohen}{1999}]{1999ApJ...520..833O}
Owocki S.~P., Cohen D.~H., 1999,
ApJ, 520, 833

\bibitem[\protect\citeauthoryear{Parker}{1955}]{1955ApJ...122..293P}
Parker E.~N., 1955,
ApJ, 122, 293 

\bibitem[\protect\citeauthoryear{Preibisch et al.}{1999}]{1999NewA....4..531P}
Preibisch T., Balega Y., Hofmann K.-H., Weigelt G., Zinnecker H., 1999,
NewA, 4, 531 

\bibitem[\protect\citeauthoryear{Preibisch et al.}{2005}]{2005ApJS..160..401P}
Preibisch T., et al., 2005,
ApJS, 160, 401 


\bibitem[\protect\citeauthoryear{Schmitt \& Liefke}{2004}]{2004A&A...417..651S}
Schmitt J.~H.~M.~M., Liefke C., 2004, 
A\&A, 417, 651 


\bibitem[\protect\citeauthoryear{Stelzer et al.}{2003}]{2003A&A...407.1067S}
Stelzer B., Hu{\'e}lamo N., Hubrig S., Zinnecker H., Micela G., 2003,
A\&A, 407, 1067 

\bibitem[\protect\citeauthoryear{Stelzer et al.}{2006a}]{2006A&A...452.1001S}
Stelzer B., Hu{\'e}lamo N., Micela G., Hubrig S., 2006a,
A\&A, 452, 1001

\bibitem[\protect\citeauthoryear{Stelzer et al.}{2006b}]{2006A&A...457..223S}
Stelzer B., Micela G., Hamaguchi K., Schmitt J.~H.~M.~M., 2006b,
A\&A, 457, 223 

\bibitem[\protect\citeauthoryear{Wallace \& Hinkle}{1997}]{1997ApJS..111..445W}
Wallace L., Hinkle K., 1997,
ApJS, 111, 445 

\bibitem[\protect\citeauthoryear{Zinnecker \& Bate}{2002}]{2002ASPC..267..209Z}
Zinnecker H., Bate M.~R., 2002,
ASPC, 267, 209 

\end{thebibliography}
\end{document}